\documentclass[11pt,epsf,psfig]{article}
 \usepackage{graphicx}
\title{Some recent (and surprising) results on interface 
and contact line depinning 
in~random~media}
\author{Jean Vannimenus   
   \\[1mm] 
Laboratoire de Physique Statistique de l'ENS,\\
       24 rue Lhomond, 75005 Paris, France\thanks
	 { Laboratoire associ\'e au CNRS et aux Universit\'es Paris~VI et
	 Paris~VII.
	 }}
\date{Version of \today}
\begin{document}

\maketitle

\begin{abstract}
 I give a brief review of results obtained recently at Ecole Normale 
on the depinning transition of interfaces and contact lines,
using a variety of approaches: 
non-local Monte Carlo algorithms, 
dynamical renormalisation group calculations to 2-loop order, 
and exact solution of an infinite-range model.

\medbreak
\noindent PACS numbers: 64.60.Lx, 05.40.+j, 05.70.Ln

\medbreak

\noindent Keywords: nonequilibrium transition; depinning; 
contact line; exact solution; Edwards-Wilkinson model; Leschhorn model.


\end{abstract}
%
%
\section{Introduction}

 The motion of interfaces in random media 
(or more generally of "elastic manifolds"),
and in particular the existence of a depinning threshold
analogous to a critical point,
is a now classic problem of statistical mechanics
which appears under various forms in many areas of 
condensed matter physics~\cite{Fisher.Collective.98}.
In spite of much progress during the last 15 years 
\cite{Nattermann.Renorm.92}
many interesting questions remain open
and I take this opportunity to present some results 
recently obtained by several groups at Ecole Normale.
 A personal motivation to talk about this topic here is that
among many domains of common interest  with Gene Stanley
this is one of the few where I can claim priority,
having studied contact lines on disordered substrates
with Y. Pomeau \cite{Pomeau.Contact.85}
before Gene's own work on interface depinning  
\cite{Stanley.Interfaces.92}
- see \cite{Stanley.Fractal_concepts.95} for relevant
references up to 1995.

The incentive to look at that problem again came initially
from experimental results obtained in our laboratory
by E. Rolley and his group  
on  the helium-cesium contact line 
\cite{Rolley.Contact.98,Prevost.Contact.02}.
 A main goal 
was to improve the determination
 of the depinning threshold in order  
to obtain more precise values for
 the roughness and velocity exponents,
in the hope 
to reconcile  numerical simulations 
with the  
experimental results  
and with new theoretical predictions 
due to LeDoussal and coworkers
\cite{Wiese.2loop.01}.
 As will become apparent from this brief review
the  results brought some 
surprises 
 and call for further work.

\section{Algorithms and simulations}

 A detailed description of the motion of a fluid-solid contact line 
is quite intricate if one tries to take into account 
in a realistic way both  the hydrodynamic aspects
\cite{Pismen.disjoining.2000}
and the substrate inhomogeneity. 
For this problem and for interface depinning in general 
most theoretical studies therefore focus on the effects of disorder 
close to the threshold, when the mean velocity 
is sufficiently small for inertial effects to be negligible,
at least as a first approximation.
 It is also customary to disregard the possibility
for the interface to develop overhangs in order to pass round
unfavorable regions. 
 With these approximations the equation of motion at zero temperature
may be written
\begin{equation}
\partial u / \partial t = K[u]  + F_{ext} + \eta (u)\; ,
\label{eqn:motion}
\end{equation}
where $u(\vec{r},t)$ is the displacement measured from a flat reference
$(d-1)$-dimensional interface.
$K[u,\vec{r}]$ 
is an elastic force that tends to keep the interface flat,
its form depends on the system considered and is non-local
for the contact line.
$F_{ext}$ is an external driving force
and $\eta (u)$ represents the  interaction with  
the disordered medium or substrate,
it will be restricted in the following to random, 
spatially uncorrelated pinning forces 
distributed according to a given probability density
$\rho(\eta)$.

An essential difference with the much studied Edwards-Wilkinson 
and Kardar-Parisi-Zhang growth models
\cite{Stanley.Fractal_concepts.95} 
is that here the disorder is quenched, as the random part of 
the force felt by an immobile interface element  remains 
constant  in time.
This non-linearity makes the problem quite difficult to tackle analytically.
Numerical simulations of (\ref{eqn:motion}) with continuous
space and time also raise technical difficulties  
\cite{Jensen.Model.95}.
In particular the interface will always get pinned by 
rare strong local defects if the  maximum distorsion 
$|u_i - u_j|$ allowed in a local move is fixed 
and if the random force is drawn from a continuous distribution, 
e.g., a Gaussian as used in field-theoretical calculations. 
Most simulations have in fact been performed on discretized
cellular-automata type  
versions of the problem,
where the transverse positions $u_i(\vec{r_i},t)$ of  $N$
interface sites are updated at regular time intervals,
the jumps being continuous or discrete. 
Such simulations are rather straightforward to perform, 
at least in low dimensions, but even for these simplified versions 
it is not easy to obtain accurate values of the depinning threshold
and of the associated critical exponents 
\cite{Leschhorn.Model.92,Jost.simul.97}.
The interface velocity becomes extremely small and fluctuates
strongly, so very long simulation times 
and large system sizes are necessary.
Another problem is that it is difficult to identify
the equation that results in the continuum limit.  

 This led Rosso and Krauth \cite{Rosso.MonteCarlo.01}
to consider non-local Monte Carlo algorithms
that bypass the dynamics and give direct access to the stationary
configurations of the continuum equation (\ref{eqn:motion}), 
in the pinned phase.
When the external force is increased the last configuration 
to get destabilized is the critical interface, it
is unique for a finite sample with perodic boundary conditions.
Rosso and Krauth showed how this interface may be efficiently
 determined using an algorithm in which one looks at each time step 
for the "front" of minimal length that it is favourable to move
by one lattice unit.
The roughness exponent $\zeta$ for the interface width $w$
as a function of its lateral extension,
$w \sim  L^{\zeta}$,
may thus be obtained right at the critical point 
with better accuracy
than through the direct dynamical approach.

 For short-range elastic interactions with bounded distorsions 
the result 
is $\zeta \simeq 0.63$ in $d=2$
\cite{Krauth.zeta.01},
 in very good agreement
with the value obtained from simulations 
on  cellular-automata models
\cite{Stanley.Interfaces.92,Tang.percolation.92}
which had led to the conjecture 
\begin{equation}
\zeta = \nu_{\perp} / \nu_{\parallel} = 0.633 \pm 0.001 ,
\label{eqn:zeta}
\end{equation}
where $\nu_{\perp}$ and $\nu_{\parallel}$
are the transverse and longitudinal exponents for directed percolation. 
On the other hand, if the local distorsions  are unconstrained
and the restoring force is linear 
(i.e., the elastic energy is purely quadratic 
with respect to these distorsions),
 giving the so-called "quenched Edwards-Wilkinson" model,
the exponent is found to be
$\zeta_{QEW} = 1.17$,
in agreement with other earlier studies
\cite{Jensen.Model.95}.

If the restoring force is non-linear,  as obtained for instance
by adding a quartic term of the form 
$(u_i - u_j)^4$ in the elastic energy of the QEW model,
one recovers  $\zeta \simeq 0.63$. 
This value appears to be robust and to correspond to 
a broad class of systems.
The surprise is that it is obtained in the absence
of an anisotropy term 
\cite{Dhar.anisotropy.95},
while the "standard" value  $\zeta \simeq 1.2$
seems to be specific to the harmonic QEW model. 
It remains to take into account such higher-order terms 
in the RG approach and to determine precisely 
the different universality classes and their extent.

\begin {figure}
\hspace*{-1cm}\centerline{\includegraphics[height=5cm]{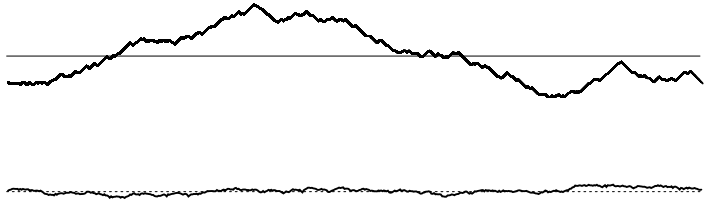}}
 \caption{
Critical interfaces  obtained using the "variant Monte Carlo"
algorithm,
 for short-range interactions with restricted height differences (top)
and for long-range interactions  (bottom).
The size is $N=512$ in both cases
(courtesy of A. Rosso).}
 \label{fig:rough}
\end {figure}

 For the contact line problem the capillary tension acting on 
the fluid surface leads to an effective long-range
interaction between the interface elements
\cite{Pomeau.Contact.85}
 and the elastic force has the form 
\begin{equation}
K[u] \propto \; \int dr' \; \frac{u(r,t) - u(r',t)}{(r - r')^2} \; .
\label{eqn:longrange}
\end{equation}
The simulations are more time-consuming in this case and
 a specific algorithm has to be devised, 
but the critical interface may still be determined accurately
\cite{Rosso.contact.01}.
It is found to be much less rough, see figure~(\ref{fig:rough}), 
with 
\begin{equation}
\zeta_{LR} = 0.388 \pm 0.002.
\label{eqn:zetaLR}
\end{equation}
This result is also surprising, as  experiments on
well-controlled  strongly disordered substrates 
\cite{Prevost.Contact.02}
give $\zeta_{exp} = 0.56 \pm 0.03 $,
much higher than initially found on
weakly disordered but ill-controlled substrates.
The calculation of $\zeta$ using dynamical-RG methods
is tricky, and the most recent work 
has shown that contributions at 2-loop order 
do not vanish 
\cite{Wiese.2loop.01},
contrary to initial expectations.
 One finds  $\zeta^{(2)} = 0.47$   at this order, 
leading to an extrapolated estimate
$\zeta_{RG} = 0.5 \pm 0.1  $,
which lies closer to the experimental results 
but further from the numerical ones than 
the one-loop value $\zeta^{(1)} = 1/3$
\cite{Kardar-Ertasz.lines.94}.

\section{An exactly solvable model}

 In the case of equilibrium phase transitions 
the critical exponents of systems with
power-law interactions such as (\ref{eqn:longrange})
are intermediate between those of short-range systems 
and the mean-field results, 
which are obtained in the limit of weak  uniform
infinite-range interactions. 
 The latter are also used as the starting point for
field-theory RG treatments and it is natural to study 
analogous models for moving interfaces
\cite{Fisher.Collective.98,Koplik.Porous.85}.
%
 For discretized models  the elastic restoring force
on site $i$  is simply given in the infinite-range limit
by  a sum over all sites
\begin{equation}
K_i[u] \propto \; \frac{1}{N} \; \sum_{j=1,N} (u_j - u_i) = \; \bar{u} - u_i,
\label{eqn:KLR}
\end{equation}
%
so the motion of an interface element is just that of
a particle in a one-dimensional potential, 
but with an additional   coupling
to the position $\bar{u}(t)$ of the center of mass,
which is to be determined self-consistently. 

 It is usually assumed in order to obtain analytical solutions
 that the infinite-range interactions have the effect
 of averaging out the 
disorder-induced fluctuations.
The interface velocity $v = d\bar{u}/dt$ 
is then independent of time in the mobile phase, in the
limit of an infinite system, 
and  obeys a self-consistent equation. 
The solution shows that 
\begin{equation}
 v \sim  \; (F_{ext} - F_c)^{\theta},
\label{vtheta}
\end{equation}
where $F_c$ is the critical force needed to depin the interface. 
For non-singular distributions of the random pinning forces
the critical velocity exponent is found to be
\cite{Fisher.Collective.98}
\begin{equation}
 \theta_{MF} = 1.
\label{thetaMF}
\end{equation}
%
 One would like to have a  rigorous proof of
this simple and physically reasonable result, 
but this is not so easy to obtain,
as it involves an interchange of limits   
when time and size go to $\infty$
which may be tricky. 
Indeed, we have been able to solve exactly a particular model in the
infinite-range limit \cite{vannimenus.interface.01}
and the exponent found in that case differs from 
the result (\ref{thetaMF}).

\subsection{The infinite-range Leschhorn model}
\label{sec:Leschhorn}

 This solvable model belongs to a family
of cellular-automaton type models due to Leschhorn 
\cite{Leschhorn.Model.92},
in which   
the interface sites
may only move forward by a discrete distance at regularly spaced
 time intervals (synchronous dynamics).
We specialize to the case where the positions $u_i$ are
integer-valued  and move by one lattice unit at a time:
\begin{equation}
u_{i}(t+1) = \left\{ \begin{array}{ll} 
              u_{i}(t) + 1  &  \mbox{\quad  if \quad $f{_i}(t) > 0, $ } \\
              u_{i}(t)      &  \mbox{\quad  if \quad $f{_i}(t) \le 0,$ }
                     \end{array} 
            \right.
\label{eqn:Lesch}
\end{equation}
where $f_i$  is the total force 
on site $i$.
 In the infinite-range limit it is given by
\begin{equation}
f_i(t) = \bar u(t) - u_i(t) + F_{ext} + g \; \eta_i(u_i),
\label{eqn:f_MF}
\end{equation}
the coupling parameter  $g$   
controls the strength of the substrate disorder,
the distribution $\rho(\eta)$ 
being normalized and of zero mean.

\begin{figure}[ht!] 
\centerline{\includegraphics[height=8cm]{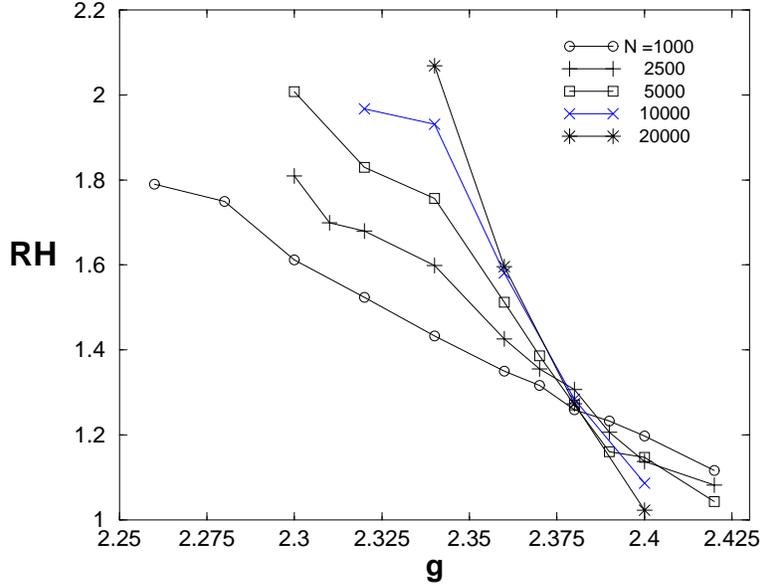}}
  \caption{ Binder ratio $R_H$ for the distribution of stopping distances 
as a function of the disorder strength $g$ 
($1000$ samples for sizes $N= 1000$ to $10000$ and 
$100$ samples of size $20000$).}
 \label{fig:ratio}
\end {figure}

 Even for such a simple system it is not straightforward to determine
 the threshold very accurately, 
as  for finite $N$ the interface only travels a finite distance $h_{m}$ 
before it stops.  
This distance increases rapidly
with $N$ in the mobile phase but it
fluctuates strongly from sample to sample close to the threshold
and the interface velocity is not precisely defined. 
%
The stopping distance plays the  same role here as the polarization
for charge-density waves
\cite{Middleton.passing.93}.
For a fixed driving force its average $<h_m>$  is expected 
to follow a scaling law of the form
\begin{equation}
 <h_m> \; \sim \; N^{\kappa} \; H[ N^y (g - g_c)] ,\label{eq:PH}
 \label{eqn:stop-distance}
\end{equation}
where the scaling function $ H(z)$ 
is finite and regular for $z = 0$,
$g_c(F_{ext})$ is the critical coupling,
and $\kappa$ and $y$ are critical exponents. 
Expression~(\ref{eqn:stop-distance})
 contains three unknown parameters and it is more 
practical and accurate to study the Binder ratio
\begin{equation}
 R_H = \;  <h_m^2>/<h_m>^2 ,
\label{eq:RH}
\end{equation}
which is expected to converge to a finite value at the critical point
when $N \to \infty$.
The results are displayed in figure~(\ref{fig:ratio})
for a symmetric uniform disorder distribution, 
$\rho(\eta)= 1/2$ for $-1 < \eta < 1$.
%
The critical ratio is $R_H^{\ast}~\simeq~1.28$,
a little below the  value $4/3$
which would correspond to a Gaussian distribution
for $h_m$,
and this approach yields an estimate
$g_c= 2.380 \pm 0.005$
for the threshold at $F_{ext} = 0$.

\subsection{Exact solution}

 An exact solution of the model may in fact be obtained 
by noting that 
deterministic evolution equations can be written down for
%
$P_{k}(x, t)$, the fraction of interface sites  at height $k$ 
which experience a local pinning force $ gx$
\cite{vannimenus.interface.01}. 
The key is to realize that due to the simple dynamics
(\ref{eqn:Lesch})
this quantity has a quasi-factorized form :
%
 \begin{equation}
 P_k(x,t)= \left\{ \begin{array}{ll}
           \lambda_{k}(t)~\rho(x) & \mbox{\quad if \quad $x < \eta_k(t)$,}\\
            \mu_{k}(t)~\rho(x)   &  \mbox{\quad if \quad $x > \eta_k(t)$,} 
            \end{array}
          \right.
 \label{eqn:distrib}
 \end{equation}
where the discontinuity point  
is related to the mean interface position by 
$\eta_{k}(t)= (k - \bar {u}(t-1))/g$.
The weights $\lambda_{k}(t)$ and $\mu_{k}(t)$ 
 obey  (relatively) simple  
recursion relations 
and the end result is a closed  system of equations
describing the interface dynamics in the thermodynamic limit.
%

In the mobile phase ($g < g_c$),
this system admits only time-dependent solutions
so the threshold may be determined by studying the existence
of stationary solutions. 
For a uniform $\rho(\eta)$
the calculation can be done analytically and
the threshold is given by the value of $g$
for which a certain polynomial has a double root.    
For small driving forces one obtains 
\cite{vannimenus.interface.01}
 \begin{equation}
g_c  =  2.38006232... +  \; 2.3901... \; F_{ext}.
 \label{eqn:gcrit}
 \end{equation}
 Note that a threshold exists even for a negative $ F_{ext}$ :
Some interface elements will move forward against the external bias,
if the local pinning force $g~\eta_i$ acting on them is strong enough, 
and due to the asymmetry of the dynamics (\ref{eqn:Lesch})
this is not compensated
by backward motions of other elements. 
This   
"ratchet" effect is of collective origin
- an individual particle in a negatively biased 
random potential would eventually stop -,
and it is driven by quenched disorder
rather than by random diffusion
as in models recently proposed for molecular motors
 \cite{Ajdari.ratchet.97}.

The threshold may also be obtained more directly
through the following  
argument. 
 Let us assume that
 in the pinning phase,
among all the possible  equilibrium positions
for  an interface element,
the one actually reached dynamically  
is the first lattice site $ z^{\ast}$ 
for which $f_i$ is non-negative 
\cite{Fisher.Collective.98}, 
i.e.:
\begin{eqnarray}
 f(z^{\ast}) & = & \bar z - z^{\ast} + g \; \eta(z^{\ast}) \quad \le 0, 
 \label{eqn:zstopping} \\
 f(z) &  = & \bar z - z + g \; \eta(z) \qquad > 0,   \qquad \mbox{for all }
 \;   z < z^{\ast},
 \label{eqn:znonstopping}
 \end{eqnarray}
where  the position $\bar z$ of the center of mass
can be restricted to $ 0 \le \bar z < 1$
due to statistical translational invariance.
The probability density of the stopping point is given by
 \begin{equation}
\mbox{$\mathcal{P}$} (z^{\ast}) = [ \prod_{z_i < z^{\ast}} R(z_i) ] \;
   [ 1 - R(z^{\ast}) ],
 \label{eqn:Pzstop}
 \end{equation}
where $R(z)$ is the probability that $(z - \bar z) < g \; \eta(z)$.
For a uniform symmetric  distribution $\rho(\eta)$ one has
\begin{equation}
 R(z) = \left\{ \begin{array}{ll} 
                 1  &  \mbox{\quad  if } \quad z < \bar z - g ,\\
                 0      &  \mbox{\quad  if  } \quad z > \bar z + g ,\\
			              \frac{1}{2} \; (1 - \frac{z- \bar z}{g})
                       & \quad \mbox{otherwise} .
                  \end{array} 
            \right.  
  \label{eqn:Rz} \\ 
\end{equation}
 The self-consistency condition on the center of mass,
$ \bar{z} = \sum  z^{\ast} 
              \mbox{$\mathcal{P}$}(z^{\ast})  $,
then yields a polynomial equation in $\bar z$
whose coefficients are functions of $g$ and the condition 
for existence of acceptable solutions
can be shown to be identical to the one obtained above 
for the threshold $g_c$.

 The exact solution also shows that the interface has a bounded width
and the average stopping distance $<h_m>$ is finite at~$g_c$,
so 
\begin{equation}
\zeta  \; = \; \kappa \; = \; 0 ,
\label{eqn:zetakappa}  
\end{equation}
in agreement with standard mean-field predictions.
An analysis of the numerical results shown in 
figure~(\ref{fig:ratio})
gives a rough estimate $y \sim 0.5$ 
for the size exponent in~(\ref{eqn:stop-distance}).

Once the threshold is known exactly
it is possible to study the interface dynamics in its 
vicinity in detail, analytically and numerically.
One finds that in contrast with the assumption usually made 
the velocity $v(t)$ is strongly non-uniform:
the interface moves through rapid bursts, separating periods
of very slow motion. 
The minimum velocity $v_{min}$ vanishes linearly
but the average velocity~$\bar v$, measured over a very long 
time interval, vanishes at the threshold as
\begin{equation}
\bar {v}     \simeq   0.06811... (g_c -g)^{1/2} ,
\label{eqn:vaverage}  
\end{equation}
which corresponds to a critical exponent 
\begin{equation}
 \theta = 1/2.
\label{thetaDMF}
\end{equation}

This result is surprising at first sight, as it differs from
the standard mean-field prediction, $\theta_{MF} = 1$.
It means that the existence of an underlying lattice
is a relevant perturbation, in the RG sense.
This can be understood physically, by noting that 
mean-field theory predicts the interface to be smooth
($\zeta = 0$), i.e., its width remains comparable to the
lattice spacing, so the continuum limit cannot be taken
naively. 
In RG terms the discreteness prevents the effective random potential
seen by the interface on large scales  
from developing a cusp at the origin~\cite{Wiese.2loop.01}.

It remains to be seen if  lattice effects also play 
a role for systems with long-range interactions such 
as the contact line.
Numerically one finds that the crossover region
 between a linear and a square-root
dependence of the form~(\ref{eqn:vaverage})
 is rather narrow,
so it would be difficult to extract the correct 
critical behaviour from simulations
such as those  
in section~(\ref{sec:Leschhorn})
- this should be kept in mind when analyzing numerical results
in finite dimensions.

\bigskip

I thank B. Derrida and A. Rosso for stimulating discussions, 
and 
 M. M\'{e}zard
for suggesting the alternative approach to derive the threshold value.


%

\end{document}